\documentclass{llncs}

\usepackage{graphicx}     
\usepackage{pst-all}      

\def\dnf{{\rm DNF}}

\def\ll{{\langle}}
\def\rl{{\rangle}}
\def\bt{{\bf t}}
\def\rules{{\Delta}}

\newcommand\sem[2]{{[\!\![#2]\!\!]}_{#1}}
\newcommand\proj[2]{{#1\!\!\downarrow\!\!#2}}
\newcommand\tempty[1]{{{\rm empty}(#1)}}
\newcommand {\definedas}{\stackrel{\rm def}{=}} 
\newcommand {\first}[1]{{\it #1}} 
\def\terms{{\cal T} } 
\newcommand\tlta[1]{{{\rm TLA}(#1)}}
\newcommand\rta[1]{{{\rm RTA}(#1)}}
\def\bt{{\bf t}}

\def\endofproof{~\hspace*{\fill}$\rule{1ex}{1.5ex}$\par}

\def\true{{\rm true} } 
\def\false{{\rm false}} 
\def\lor{\vee} 
\def\land{\wedge} 

\def\cleq{\preceq}

\def\ttop{{\bf 1} } 
\def\tbot{{\bf 0} } 
\def\tand{{\sqcap} } 
\def\tor{{\sqcup} } 
\def\bigtand{{\sqcap} } 
\def\bigtor{{\sqcup} } 
\def\tneg{{\sim} } 
\def\fcal{{\cal F}}
\def\acal{{\cal A}}
\def\bcal{{\cal B}}
\def\ocal{{\cal O}}
\def\mcal{{\cal M}}
\def\ncal{{\cal N}}

\def\asf{{\sf A}}
\def\bsf{{\sf B}}
\def\csf{{\sf C}}

\newcommand\comments[1]{}

\begin{document} 

\large

\bibliographystyle{plain}

\title{An Emptiness Algorithm for Regular Types with Set Operators} 
\author{Lunjin Lu and John G. Cleary}
\institute{ Department of Computer Science\\ University of Waikato \\
Hamilton, New Zealand\\ Phone: +64-838-4627/4378\\
\{lunjin,jcleary\}@cs.waikato.ac.nz}

\date{}

\maketitle
 
\begin{abstract}  
An algorithm to decide the emptiness of a regular type expression with
set operators given a set of parameterised type definitions is
presented.  The algorithm can also be used to decide the equivalence
of two regular type expressions and the inclusion of one regular type
expression in another.  The algorithm strictly generalises previous
work in that tuple distributivity is not assumed and set operators are
permitted in type expressions.

\end{abstract}

\section{Introduction}
Types play an important role in programming
languages~\cite{Cardelli:Wegner:85}. They make programs easier to
understand and help detect errors.  Types have been introduced into
logic programming in the forms of type checking and
inference~\cite{Beierle:ICLP95,Dart:Zobel:JLP92,FruhwirthSVY:LICS91,Mycroft:OKeefe:84,Yardeni:Shapiro:91}
or type
analysis~\cite{Mishra:84,Zobel:87,HeintzeJ90,HeintzeJ92,GallagherW94,Lu95,CL96,LuJLP98}
or typed
languages~\cite{Hanus:TCS,Jacods:PLDI90,Reddy:NACLP90,YardeniFS:ICLP91}.
Recent logic programming systems allow the programmer to declare types
for predicates and type errors are then detected either at compile
time or at run time. \comments{Even in early logic programming
systems, built-in predicates are usually typed and type checking for
these predicates are performed at run time.}  The reader is referred
to \cite{Pfenning92} for more details on types in logic programming.

A \first{type} is a possibly infinite set of ground terms with a
finite representation. An integral part of any type system is its type
language that specifies which sets of ground terms are types. To be
useful, types should be closed under intersection, union and
complement operations. The decision problems such as the emptiness of
a type, inclusion of a type in another and equivalence of two types
should be decidable.  Regular term
languages~\cite{GecsegS84,ComonDGLTT98}, called regular types, satisfy
these conditions and have been used widely used as
types~\cite{Soloman78,Mishra:84,Zobel:87,Dart:Zobel:JLP92,HeintzeJ90,Jacods:PLDI90,Reddy:NACLP90,YardeniFS:ICLP91,FruhwirthSVY:LICS91,Yardeni:Shapiro:91,HeintzeJ92,GallagherW94,Lu95,CL96,LuJLP98}.

Most type systems use \first{tuple distributive regular types} which
are strictly less powerful than regular
types~\cite{Soloman78,Mishra:84,Zobel:87,HeintzeJ90,Jacods:PLDI90,Reddy:NACLP90,YardeniFS:ICLP91,FruhwirthSVY:LICS91,Yardeni:Shapiro:91,HeintzeJ92,GallagherW94,Lu95,CL96,LuJLP98}. Tuple
distributive regular types are regular types closed under tuple
distributive closure. Intuitively, the tuple distributive closure of a
set of terms is the set of all terms constructed recursively by
permuting each argument position among all terms that have the same
function symbol~\cite{Yardeni:Shapiro:91}.

This paper gives an algorithm to decide if a type expression denotes
an empty set of terms. The correctness of the algorithm is proved and
its complexity is analysed.  The algorithm works on prescriptive
types~\cite{Reddy:NACLP90}.  By prescriptive types, we mean that the
meaning of a type is determined by a given set of type definitions.
We allow parametric and overloading polymorphism in type
definitions. Prescriptive types are useful both in compilers and other
program manipulation tools such as debuggers because they are easy to
understand for programmers. Type expressions may contain set operators
with their usual interpretations. Thus, the algorithm can be used to
decide the equivalence of two type expressions and the inclusion of
one type expression in another.  The introduction of set operators
into type expressions allows concise and intuitive representation of
regular types.

Though using regular term languages as types allow us to make use of
theoretical results in the field of tree automata~\cite{GecsegS84},
algorithms for testing the emptiness of tree automata cannot be
applied directly as type definitions may be parameterised. For
instance, in order to decide the emptiness of a type expression given
a set of type definitions, it would be necessary to construct a tree
automaton from the type expression and the set of type definitions
before an algorithm for determining the emptiness of an tree automaton
can be used. When type definitions are parameterised, this would make
it necessary to construct a different automaton each time the
emptiness of a type expression is tested. Thus, an algorithm that
works directly with type definitions is desirable as it avoids this
repeated construction of automata.

 Attempts have been made in the past to find algorithms for
regular
types~\cite{Mishra:84,FruhwirthSVY:LICS91,Yardeni:Shapiro:91,Zobel:87,YardeniFS:ICLP91,DartZ92,Dart:Zobel:JLP92}.
To our knowledge, Dart and Zobel's work~\cite{DartZ92} is the only one
to present decision algorithms for emptiness and inclusion problems
for prescriptive regular types without the tuple distributive
restriction. Unfortunately, their decision algorithm for the inclusion
problem is incorrect for regular types in general. See~\cite{LuC:note}
for a counterexample. Moreover, the type language of Dart and Zobel is
less expressive than that considered in this paper since it doesn't
allow set operators and parameterised type definitions. 

Set constraint solving has also been used in type checking and type
inference~\cite{AikenW92,AikenL94,heintze-set,heintze-decision,DevienneTT98}.
However, set constraint solving methods are intended to infer
descriptive types~\cite{Reddy:NACLP90} rather than for testing
emptiness of prescriptive types~\cite{Reddy:NACLP90}. Therefore, they
are useful in different settings from the algorithm presented in this
paper. Moreover, algorithms proposed for set constraint
solving~\cite{AikenW92,AW93,AikenL94,AikenKVW93} are not applicable to
the emptiness problem we considered in this paper as they don't take
type definitions into account.

The remainder of this paper is organised as
follows. Section~\ref{sec:regular} describes our language of type
expressions and type definitions. Section~\ref{sec:algorithm} presents
our algorithm for testing if a type expression denotes an empty set of
terms.  Section~\ref{sec:soundness} addresses the of the algorithm.
Section~\ref{sec:complexity} presents the complexity of the algorithm
and section~\ref{sec:conclusion} concludes the paper. Some lemmas are
presented in the appendix.

\section{Type Language} \label{sec:regular}

      Let $\Sigma$ be a fixed ranked alphabet. Each symbol in $\Sigma$
      is called a function symbol and has a fixed arity. It is assumed
      that $\Sigma$ contains at least one constant that is a function
      symbol of arity $0$. The arity of a symbol $f$ is denoted as
      $arity(f)$. $\Sigma$ may be considered as the set of function
      symbols in a program.  Let $\terms(\Phi)$ be the set of all
      terms over $\Phi$. $\terms(\Sigma)$ is the set of all possible
      values that a program variable can take. We shall use regular
      term languages over $\Sigma$ as types.

      A type is represented by a ground term constructed from another
      ranked alphabet $\Pi$ and $\{\tand,\tor,\tneg,\ttop,\tbot\}$,
      called type constructors. It is assumed that
      $(\Pi\cup\{\tand,\tor,\tneg,\ttop,\tbot\})\cap\Sigma=\emptyset$. Thus,
      a type expression is a term in
      $\terms(\Pi\cup\{\tand,\tor,\tneg,\ttop,\tbot\})$.  The
      denotations of type constructors in $\Pi$ are determined by type
      definitions whilst $\tand,\tor,\tneg,\ttop$ and $\tbot$ have
      fixed denotations that will be given soon.

      Several equivalent formalisms such as tree
      automata~\cite{GecsegS84,ComonDGLTT98}, regular term
      grammars~\cite{GecsegS84,DartZ92,ComonDGLTT98} and regular unary
      logic programs~\cite{Yardeni:Shapiro:91} have been used to
      define regular types. We define types by type rules. A type
      rule is a production rule of the form
      $c(\zeta_1,\cdots,\zeta_{m})\rightarrow \tau$ where $c\in\Pi$,
      $\zeta_1,\cdots,\zeta_{m}$ are different type parameters and
      $\tau\in \terms(\Sigma\cup\Pi\cup\Xi_{m})$ where
      $\Xi_{m}=\{\zeta_1,\cdots,\zeta_{m}\}$.  The restriction that
      every type parameter in the righthand side of a type rule must
      occur in the lefthand side of the type rule is often referred to
      as {type preserving}~\cite{Tiuryn:90} and has been used in all
      the type definition formalisms. Note that overloading of
      function symbols is permitted as a function symbol can appear in
      the righthand sides of many type rules. We denote by $\rules$
      the set of all type rules and define
      $\Xi\definedas\bigcup_{c\in\Pi}\Xi_{arity(c)}$. $\ll\Pi,\Sigma,\rules\rl$
      is a restricted form of context-free term grammar.

\begin{example} \label{ex0} Let $\Sigma=\{0,s(),nil,cons(,)\}$ and 
                    $\Pi=\{Nat,Even,List()\}$.
                $\rules$ defines natural numbers,
                even numbers, and lists where
\[ \rules= \left\{\begin{array}{l}
                   Nat \rightarrow 0 ~|~ s(Nat),\\
                   Even \rightarrow 0 ~|~ s(s(Even)),\\
                   List(\zeta) \rightarrow nil~|~ cons(\zeta,List(\zeta))
                  \end{array}
           \right\}
\]
where, for instance, $ Nat \rightarrow 0 ~|~ s(Nat)$ is an abbreviation
of two rules $Nat \rightarrow 0$ and $Nat \rightarrow s(Nat)$.

\endofproof
\end{example}

$\rules$ is called simplified if $\tau$ in each production rule
$c(\zeta_1,\cdots,\zeta_{m})\rightarrow \tau$ is of the form
$f(\tau_1,\cdots,\tau_{n})$ such that each $\tau_j$, for
$1\leq{j}\leq{n}$, is either in $\Xi_{m}$ or of the form
$d(\zeta_1',\cdots,\zeta_k')$ and
$\zeta_1',\cdots,\zeta_k'\in\Xi_{m}$.  We shall assume that $\rules$
is simplified. There is no loss of generality to use a simplified set
of type rules since every set of type rules can be simplified by
introducing new type constructors and rewriting and adding type rules
in the spirit of \cite{DartZ92}.

\begin{example} \label{ex1} The following 
is the simplified version of the set of type rules in
example~\ref{ex0}.  $\Sigma=\{0,s(),nil,cons(,)\}$,
$\Pi=\{Nat,Even,Odd, List()\}$ and
\[ \rules= \left\{\begin{array}{ll}
                   Nat \rightarrow 0~|~ s(Nat), & 
                   Even \rightarrow 0~|~ s(Odd),\\
                   Odd\rightarrow s(Even), &
                   List(\zeta) \rightarrow nil~|~ cons(\zeta,List(\zeta)) 
                  \end{array}
           \right\}
\]

\endofproof
\end{example}

A type valuation $\phi$ is a mapping from $\Xi$ to
$\terms(\Pi\cup\{\tand,\tor,\tneg,\ttop,\tbot\})$.  The instance
$\phi(R)$ of a production rule $R$ under $\phi$ is obtained by
replacing each occurrence of each type parameter $\zeta$ in $R$ with
$\phi(\zeta)$.  E.g., $List(Nat\tand(\tneg Even)) \rightarrow
cons(Nat\tand(\tneg Even),List(Nat\tand(\tneg Even)))$ is the instance
of $List(\zeta) \rightarrow cons(\zeta,List(\zeta))$ under a type
valuation that maps $\zeta$ to $Nat\tand(\tneg Even)$.  Let
\begin{eqnarray*} {ground}(\rules) & \definedas& 
 \{\phi(R)~|~R\in\rules\land\phi\in(\Xi\mapsto\terms(\Pi\cup\{\tand,\tor,\tneg,\ttop,\tbot\}))\}\\
                  & & \cup ~~\{\ttop\mapsto f(\ttop,\cdots,\ttop)~|~f\in\Sigma\}
\end{eqnarray*}
${ground}(\rules)$ is the set of all ground instances of grammar rules
in $\rules$ plus rules of the form $\ttop\rightarrow
f(\ttop,\cdots,\ttop)$ for every $f\in\Sigma$.

Given a set $\rules$ of type definitions, the
type denoted by a type expression is determined by the following
meaning function.
\begin{eqnarray*}
\sem{\rules}{\ttop} &\definedas & \terms(\Sigma)\\
\sem{\rules}{\tbot} &\definedas & \emptyset\\
\sem{\rules}{E_1\tand E_2} &\definedas & \sem{\rules}{E_1}\cap\sem{\rules}{E_2}\\
\sem{\rules}{E_1\tor E_2} &\definedas & \sem{\rules}{E_1}\cup\sem{\rules}{E_2}\\
\sem{\rules}{\tneg E} &\definedas & \terms(\Sigma)- \sem{\rules}{E}\\
\sem{\rules}{\omega} &\definedas & \bigcup_{(\omega\rightarrow f(E_1,\cdots,E_{n}))\in{ground}(\rules)} 
         \{f(t_1,\cdots,t_{n})~|~\forall 1\leq{i}\leq{n}.~
                                        t_{i}\in\sem{\rules}{E_{i}}\}
\end{eqnarray*}

$\sem{\rules}{\cdot}$ gives fixed denotations to $\tand,\tor,\tneg,\ttop$ and
$\tbot$. \comments{$\sem{\rules}{\ttop}=\terms(\Sigma)$ as $\ttop\rightarrow
f(\ttop,\cdots,\ttop)$ is in $ground(\rules)$. $\sem{\rules}{\tbot}=\emptyset$
as there is no $E_1,\cdots,E_n$ such that $\tbot\rightarrow
f(E_1,\cdots,E_n)$ is in $ground(\rules)$.} $\tand$, $\tor$ and
$\tneg$ are interpreted by $\sem{\rules}{\cdot}$ as set intersection, set
union and set complement with respect to $\terms(\Sigma)$. $\ttop$
denotes $\terms(\Sigma)$ and $\tbot$ the empty set.

\begin{example} Let $\rules$ be that  in example~\ref{ex1}. We have 
\begin{eqnarray*}
 \sem{\rules}{Nat} &= &  \{0,s(0),s(s(0)),\cdots\}\\
 \sem{\rules}{Even} &=& \{0,s(s(0)),s(s(s(s(0)))),\cdots\}\\
 \sem{\rules}{Nat\tand\tneg Even} &=& \{s(0),s(s(s(0))),s(s(s(s(s(0))))),
                                  \cdots\}\\ 
 \sem{\rules}{List(Nat\tand\tneg Even)} &=&
        \{cons(s(0),nil),cons(s(s(s(0))),nil),\cdots\}
\end{eqnarray*} 
\end{example}  

The lemma~\ref{lm:regular} in the appendix states that {\it every type
expression denotes a regular term language}, that is, a regular type.

We extend $\sem{\rules}{\cdot}$ to sequences $\theta$ of type expressions as
follows.
\begin{eqnarray*} 
\sem{\rules}{\epsilon} & \definedas &\{\epsilon\}\\
\sem{\rules}{\ll{E}\rl\bullet\theta'} & \definedas & 
                    \sem{\rules}{E}\times\sem{\rules}{\theta'}
\end{eqnarray*}
where $\epsilon$ is the empty sequence, $\bullet$ is the infix
sequence concatenation operator, $\ll{E}\rl$ is the sequence
consisting of the type expression $E$ and $\times$ is the Cartesian
product operator. As a sequence of type expressions, $\epsilon$ can be
thought of consisting of zero instance of $\ttop$. We use $\Lambda$ to
denote the sequence consisting of zero instance of $\tbot$ and define
$\sem{\rules}{\Lambda}=\emptyset$.

We shall call a sequence of type expressions simply a sequence. A
sequence expression is an expression consisting of sequences of
the same length and $\tand$, $\tor$ and $\tneg$.  The length of
the sequences in a sequence expression $\theta$ is called the
dimension of $\theta$ and is denoted by $\|\theta\|$.  Let
$\theta,\theta_1$ and $\theta_2$ be sequence expressions of the
same length.
\begin{eqnarray*} 
\sem{\rules}{\theta_1\tand\theta_2} &\definedas&
                   \sem{\rules}{\theta_1}\cap\sem{\rules}{\theta_2}\\
                   \sem{\rules}{\theta_1\tor\theta_2} &\definedas&
                   \sem{\rules}{\theta_1}\cup\sem{\rules}{\theta_2}\\
                   \sem{\rules}{\tneg\theta} &\definedas &
                   \underbrace{\terms(\Sigma)\times\cdots\times\terms(\Sigma)}_{\mbox{$\|\theta\|$ times}}-\sem{\rules}{\theta}
\end{eqnarray*}
A conjunctive sequence expression is a sequence expression of the
form $\gamma_1\land\cdots\land\gamma_{m}$ where $\gamma_i$ for,
$1\leq{i}\leq{m}$, are sequences.

\section{Emptiness Algorithm} \label{sec:algorithm}
This section presents an algorithm that decides if a type expression
denotes the empty set with respect to a given set of type
definitions. The algorithm can also be used to decide if (the
denotation of) one type expression is included in (the denotation of)
another because $E_1$ is included in $E_2$ iff $E_1\tand\tneg E_2$ is
empty.

We first introduce some terminology and notations. A type atom is a
type expression of which the principal type constructor is not a set
operator.  A type literal is either a type atom or the complement of a
type atom. A conjunctive type expression $C$ is of the form
$\tand_{i\in{I}}{\bf l}_{i}$ with ${\bf l}_{i}$ being a type
literal. Let $\alpha$ be a type atom. $\fcal(\alpha)$ defined below is
the set of the principal function symbols of the terms in
$\sem{\rules}{\alpha}$.
\[\fcal(\alpha)\definedas\{f\in\Sigma~|~\exists\zeta_1\cdots\zeta_{k}.
((\alpha\rightarrow f(\zeta_1,\cdots,\zeta_{k}))\in{ground}(\rules))\}
\]  
Let $f\in\Sigma$. Define
\[\acal_{\alpha}^{f} \definedas  \{\ll\alpha_1,\cdots,\alpha_{k}\rl~|~
             (\alpha\rightarrow
             f(\alpha_1,\cdots,\alpha_{k}))\in{ground}(\rules)\}
\]
We have
$\sem{\rules}{\acal_{\alpha}^{f}}=\{\ll{t}_1,\cdots,{t}_{k}\rl~|~
f({t}_1,\cdots,{t}_{k})\in\sem{\rules}{\alpha}\}$.  Both
$\fcal(\alpha)$ and $\acal_{\alpha}^{f}$ are finite even though
${ground}(\rules))$ is usually not finite. \comments{An implementation
can use unification rather than set membership to compute
$\fcal(\alpha)$ and $acal_{\alpha}^{f}$.}

The algorithm repeatedly reduces the emptiness problem of a type
expression to the emptiness problems of sequence expressions and
then reduces the emptiness problem of a sequence expression to
the emptiness problems of type expressions. Tabulation is used to
break down any possible loop and to ensure termination.  Let $O$
be a type expression or a sequence expression. Define
$\tempty{O}\definedas (\sem{\rules}{O}=\emptyset)$.

\subsection{Two Reduction Rules}
We shall first sketch the two reduction rules and then add tabulation
to form an algorithm.  Initially the algorithm is to decide the
validity of a formula of the  form
\begin{equation} \label{fm:type}
\tempty{E}
\end{equation} where $E$ is a type expression. 

\subsubsection{Reduction Rule One.} 
The first reduction rule rewrites a formula of the
      form~(\ref{fm:type}) into a conjunction of formulae of the
      following form.  \begin{equation} \label{fm:sequence}
      \tempty{\sigma} \end{equation} where $\sigma$ is a sequence
      expression where $\tneg$ is applied to type expressions but not
      to any sequence expression.

It is obvious that a type expression has a unique (modulo
equivalence of denotation) disjunctive normal form. Let ${\dnf}(E)$
be the disjunctive normal form of $E$. $\tempty{E}$ can written
into $\land_{C\in{\dnf}(E)}\tempty{C}$. Each $C$ is a conjunctive
type expression.  We assume that $C$ contains at least one
positive type literal. This doesn't cause any loss of generality
as $\sem{\rules}{\ttop\tand C}=\sem{\rules}{C}$ for any conjunctive type
expression $C$. We also assume that $C$ doesn't contain repeated
occurrences of the same type literal.

Let $C=\bigtand_{1\leq{i}\leq{m}}\omega_{i} ~\tand~
             \bigtand_{1\leq{j}\leq{n}}\tneg \tau_{j}$ where
             $\omega_{i}$ and $\tau_{j}$ are type atoms. The set of
             positive type literals in $C$ is denoted as
             $pos(C)\definedas\{\omega_{i}~|~1\leq{i}\leq{m}\}$ while
             the set of complemented type atoms are denoted as
             $neg(C)\definedas\{\tau_{j}~|~1\leq{j}\leq{n}\}$.
             $lit(C)$ denotes the set of literals occurring in $C$. By
             lemma~\ref{lm:x} in the appendix, $\tempty{C}$ is
             equivalent to
\begin{equation}\begin{array}{l}
 \forall{f\in\cap_{\alpha\in{pos}(C)}\fcal(\alpha)}.\\
   ~~\tempty{(\bigtand_{\omega\in{pos}(C)}(\tor \acal_{\omega}^f))
   \tand (\bigtand_{\tau\in{neg}(C)}\tneg(\tor \acal_{\tau}^f))}
   \end{array} 
\label{eq:star}
\end{equation}

The intuition behind the equivalence is as follows.  $\sem{\rules}{C}$
 is empty iff, for every function symbol $f$, the set of the sequences
 $\ll t_1,\cdots, t_{k}\rl$ of terms such that $f(t_1,\cdots,
 t_{k})\in\sem{\rules}{C}$ is empty. Only the function symbols
 in $\cap_{\alpha\in{pos}(C)}\fcal(\alpha)$ need to be considered.

We note the following two special cases of the  formula (\ref{eq:star}).
\begin{itemize}
\item [(a)] If $\cap_{\alpha\in{pos}(C)}\fcal(\alpha)=\emptyset$ then
      the formula (\ref{eq:star}) is true because $\land\emptyset=\true$. In
      particular, $\fcal(\tbot)=\emptyset$. Thus, if
      $\tbot\in{pos}(C)$ then
      $\cap_{\alpha\in{pos}(C)}\fcal(\alpha)=\emptyset$ and hence the
      formula (\ref{eq:star}) is true.
\item [(b)] If $\acal_{\tau}^f=\emptyset$ for some $\tau\in{neg}(C)$
      then $\tor \acal_{\tau}^f=\ll\tbot,\cdots,\tbot\rl$ and
      $\tneg(\tor \acal_{\tau}^f)=\ll\ttop,\cdots,\ttop\rl$. Thus,
      $\tau$ has no effect on the subformula for $f$ when
      $\acal_{\tau}^f=\emptyset$.
\end{itemize}

In order to get rid of complement operators over sequence
sub-expressions, the complement operator in $\tneg(\tor
\acal_{\tau}^f)$ is pushed inwards \comments{ by making use of the
following two properties of sequence expressions.
\begin{itemize} 
\item $
\sem{\rules}{\tneg\bigtor_{i\in{I}}\sigma_{i}} =
\sem{\rules}{\bigtand_{i\in{I}}\tneg\sigma_{i}}$, and 
\item $\sem{\rules}{\tneg\ll E_1,E_2,\cdots,E_{k}\rl} =
\sem{\rules}{\bigtor_{1\leq{l}\leq{k}}
\ll\underbrace{\ttop,\cdots,\ttop}_{l-1},\tneg
E_{l},\underbrace{\ttop,\cdots,\ttop}_{k-l}\rl}$.  \end{itemize}
This is done} 
by the function $push$ defined in the following.
\begin{eqnarray*} 
push(\tneg (\tor_{i\in{I}}\gamma_{i})) &\definedas&
               \bigtand_{i\in{I}}push(\tneg\gamma_{i})\\
push(\tneg\ll E_1,E_2,\cdots,E_{k}\rl) &\definedas & 
    \bigtor_{1\leq{l}\leq{k}}
\ll\underbrace{\ttop,\cdots,\ttop}_{l-1},\tneg
E_{l},\underbrace{\ttop,\cdots,\ttop}_{k-l}\rl~~~~\mbox{for $k\geq 1$}\\
push(\tneg\epsilon) &\definedas & \Lambda
\end{eqnarray*}

It follows from De Morgan's law and the definition of
$\sem{\rules}{\cdot}$ that $\sem{\rules}{push(\tneg(\tor
\acal_{\tau}^f))}=\sem{\rules}{\tneg(\tor
\acal_{\tau}^f)}$. Substituting $push(\tneg(\tor \acal_{\tau}^f))$ for
$\tneg(\tor \acal_{\tau}^f)$ in the formula (\ref{eq:star}) gives rise
to a formula of the form~(\ref{fm:sequence}).

\subsubsection{Reduction Rule Two.} 
      The second reduction rule rewrites a formula of the
      form~\ref{fm:sequence} to a conjunction of disjunctions of
      formulae of the form~\ref{fm:type}.
      Formula~\ref{fm:sequence} is written into a disjunction of
      formulae of the  form.  \[ \tempty{\Gamma} \] where
      $\Gamma$ be a conjunctive sequence expression. 
      
      In the case $\|\Gamma\|=0$, by lemma~\ref{lm:xxx} in the
      appendix, $\tempty{\Gamma}$ can be decided without further
      reduction. If $\Lambda\in\Gamma$ then $\tempty{\Gamma}$ is true
      because $\sem{\rules}{\Lambda}=\emptyset$.  Otherwise,
      $\tempty{\Gamma}$ is false because
      $\sem{\rules}{\Gamma}=\{\epsilon\}$.

      In the case $\|\Gamma\|\neq 0$, $\tempty{\Gamma}$ is equivalent
      to \[\lor_{1\leq{j}\leq\|\Gamma\|} \tempty{\proj{\Gamma}{j}}\]
      where, letting $\Gamma=\gamma_1\tand\cdots\tand\gamma_k$,
      $\proj{\Gamma}{j}\definedas
      \bigtand_{1\leq{i}\leq{k}}\gamma_{i}^{j}$ with $\gamma_i^j$
      being the $j^{th}$ component of $\gamma_{i}$. Note that
      $\proj{\Gamma}{j}$ is a type expression and
      $\tempty{\proj{\Gamma}{j}}$ is of the form~\ref{fm:type}.

\subsection{Algorithm} 

The two reduction rules in the previous section form the
core of the algorithm. However, they alone cannot be used as an
algorithm as a formula $\tempty{E}$ may reduce to a formula
containing $\tempty{E}$ as a sub-formula, leading to
nontermination. Suppose $\Sigma=\{f(),a\}$, $\Pi=\{Null\}$ and
$\rules=\{Null\rightarrow f(Null)\}$. Clearly, $\tempty{Null}$ is
true. However, by the first reduction rule, $\tempty{Null}$
reduces to $\tempty{\ll Null\rl}$ which then reduces to
$\tempty{Null}$ by the second reduction rule. This process will not
terminate.

The solution, inspired by~\cite{DartZ92}, is to remember in a table a
particular kind of formulae of which truth is being tested.  When a
formula of that kind is tested, the table is first looked up. If the
formula is implied by any formula in the table, then it is determined
as true. Otherwise, the formula is added into the table and then
reduced by a reduction rule.

The emptiness algorithm presented below remembers every conjunctive
type expression of which emptiness is being tested. Thus the table is
a set of conjunctive type expressions.  Let $C_1$ and $C_2$ be
conjunctive type expressions. We define $(C_1\cleq C_2)\definedas
(lit(C_1)\supseteq lit(C_2))$. Since
$C_{i}=\bigtand_{{\bf l}\in{lit}(C_{i})}{\bf l}$, $C_1\cleq C_2$
implies $\sem{\rules}{C_1}\subseteq\sem{\rules}{C_2}$ and hence
$(C_1\cleq C_2)\land\tempty{C_2}$ implies $\tempty{C_1}$.

Adding tabulation to the two reduction rules, we obtain the following
algorithm for testing the emptiness of prescriptive regular types.
Let\\ $\bcal_{C}^f=(\bigtand_{\omega\in{pos}(C)}(\tor \acal_{\omega}^f))
\tand (\bigtand_{\tau\in{neg}(C)}push(\tneg(\tor \acal_{\tau}^f)))$.

{\small
\begin{eqnarray}\label{eq:top}
etype(E)  &\definedas &  etype(E,\emptyset)\\
etype(E,\Psi)  &\definedas& \forall C\in{\dnf}(E). 
                      etype\_conj(C,\Psi)
\label{eq:rdo:1}
\end{eqnarray}
\begin{equation}\label{eq:rdo:2}
\begin{array}{l}
etype\_conj(C,\Psi)  \definedas \\
\left\{
\begin{array}{lr}
\true, & \mbox{if $pos(C)\cap{neg}(C)\neq\emptyset$,}\\
\true, & \mbox{if $\exists C'\in\Psi.C\cleq C'$,}\\
{\begin{array}{l}
 \forall{f\in\cap_{\alpha\in{pos}(C)}\fcal(\alpha)}.
    eseq(\bcal_{C}^f,\Psi\cup\{C\}),
 \end{array}}
& \mbox{otherwise.}
\end{array}\right.
\end{array}
\end{equation}
\begin{eqnarray} 
eseq(\Theta,\Psi)  & \definedas & 
\forall\Gamma\in{\dnf}(\Theta).eseq\_conj(\Gamma,\Psi)\label{eq:rdt:1}
\\
eseq\_conj(\Gamma,\Psi) &\definedas&
\left\{ \begin{array}{lr}
\true & \mbox{if $\|\Gamma\|=0\land\Lambda\in\Gamma$},\\
\false & \mbox{if $\|\Gamma\|=0\land\Lambda\not\in\Gamma$},\\
\exists 1\leq{j}\leq\|\Gamma\|. etype(\proj{\Gamma}{j},\Psi)      
& \mbox{if $\|\Gamma\|\neq 0$.}
\end{array}\right.
\label{eq:rdt:2}
\end{eqnarray}       
}

Equation~\ref{eq:top} initialises the table to the empty set.
Equations~\ref{eq:rdo:1} and~\ref{eq:rdo:2} implement the first
reduction rule while equations~\ref{eq:rdt:1} and~\ref{eq:rdt:2}
implement the second reduction rule. $etype(,)$ and $etype\_conj(,)$
test the emptiness of an arbitrary type expression and that of a
conjunctive type expression respectively.  $eseq(,)$ tests emptiness
of a sequence expression consisting of sequences and $\tand$ and
$\tor$ operators while $eseq\_conj(,)$ tests the emptiness of a
conjunctive sequence expression. The expression of which emptiness is
to be tested is passed as the first argument to these functions. The
table is passed as the second argument. It is used in $etype\_conj(,)$
to detect a conjunctive type expression of which emptiness is implied
by the emptiness of a tabled conjunctive type expression. As we shall
show later, this ensures the termination of the algorithm. Each of the
four binary functions returns true iff the emptiness of the first
argument is implied by the second argument and the set of type
definitions.

Tabling any other kind of expressions such as arbitrary type
expressions can also ensure termination. However, tabling conjunctive
type expressions makes it easier to detect the implication of the
emptiness of one expression by that of another because $lit(C)$ can be
easily computed given a conjunctive type expression $C$. In an
implementation, a conjunctive type expression $C$ in the table can be
represented as $lit(C)$. \comments{Tabling conjunctive sequence expressions
consumes more space than tabling conjunctive type expressions. Tabling
arbitrary type expression or arbitrary sequence expressions requires a
canonical representation of these expressions as well as consumes
more space than tabling conjunctive type expressions.}

The first two definitions for $etype\_conj(C,\Psi)$ in
equation~\ref{eq:rdo:2} terminates the algorithm when the
emptiness of $C$ can be decided by $C$ and $\Psi$ without using
type definitions. The first definition also excludes from the
table any conjunctive type expression that contains both a type
atom and its complement. 

\subsection{Examples} 
We now illustrate the algorithm with some examples. 

\begin{example} 
Let type definitions be given as in example~\ref{ex1}. The tree in
figure~\ref{fig:1} depicts the evaluation of $etype(Nat\tand\tneg
Even\tand\tneg Odd)$ by the algorithm. Nodes are labeled with function
calls. We will identity a node with its label.  Arcs from a node to
its children are labeled with the number of the equation that is used
to evaluate the node.  Abbreviations used in the labels are defined in
the legend to the right of the tree. Though
$\sem{\rules}{\asf}=\sem{\rules}{\bsf}$, $\asf$ and $\bsf$ are
syntactically different type expressions. The evaluation returns true,
verifying $\sem{\rules}{Nat\tand\tneg Even\tand\tneg
Odd}=\emptyset$. Consider $etype\_conj(\bsf,\{\asf\})$. We have
$\bsf\cleq \asf$ as $lit(\asf)=lit(\bsf)$. Thus, by
equation~\ref{eq:rdo:2}, $etype\_conj(\bsf,\{\asf\})=\true$.

\begin{figure}
\psset{labelsep=2pt,tnpos=a,radius=2pt}
{
\begin{minipage}[t]{.6\textwidth}
\parbox[b]{\textwidth}{
\pstree[nodesep=2pt,levelsep=25pt]{\TR{$etype(\asf)$}}
{ \pstree{\TR{$etype(\asf,\emptyset)$}\trput{3}}
  { \pstree{\TR{$etype\_conj(\asf,\emptyset)$}~[tnpos=b]{$\land$}\trput{4}}
    { \pstree{\TR{$eseq(\epsilon\tand\Lambda,\{\asf\})$}\tlput{5}}
      {\pstree{\TR{$eseq\_conj(\epsilon\tand\Lambda,\{\asf\})$}\trput{6}}
       {\TR{\true}\trput{7}}
      }
      \pstree{\TR{$eseq(\csf,\{\asf\})$}\trput{5}}
      {\pstree{\TR{$eseq\_conj(\csf,\{\asf\})$}\trput{6}}
       {\pstree{\TR{$etype(\bsf,\{\asf\})$}\trput{7}}
        {\pstree{\TR{$etype\_conj(\bsf,\{\asf\})$}\trput{4}}
         {\TR{\true}\trput{5}}
        }
       }
      }
    }
  }
}
}
\end{minipage}
\begin{minipage}[t]{.3\textwidth}
\begin{tabbing}
XX \= XXXXX\kill
Legend:\\
 $\asf =Nat\tand\tneg Even\tand\tneg Odd$\\
 $\bsf =Nat\tand\tneg Odd\tand\tneg Even$\\
 $\csf =\ll Nat\rl \tand \ll\tneg Odd\rl \tand \ll\tneg Even\rl $
\end{tabbing}
\end{minipage}
}

\caption{\label{fig:1} Evaluation of $etype(Nat\tand\tneg{Even}\tand\tneg{Odd}))$}
\end{figure}

\endofproof
\end{example}

\begin{example} 
Let type definitions be given as in example~\ref{ex1}. The tree in
figure~\ref{fig:2} depicts the evaluation of
$etype(List(Even\tand\tneg Nat))$ by the algorithm.  The
evaluation returns false, verifying $\sem{\rules}{List(Even\tand\tneg
Nat)}\neq\emptyset$. Indeed, $\sem{\rules}{List(Even\tand\tneg
Nat)}=\{nil\}$.  The rightmost node is not evaluated as its
sibling returns $\false$, which is enough to establish the falsity
of their parent node.

\begin{figure}
\psset{labelsep=2pt,tnpos=a,radius=2pt}
\begin{minipage}[t]{.6\textwidth}
\parbox[b]{\textwidth}{
\pstree[nodesep=2pt,levelsep=25pt]{\TR{$etype(\asf)$}}
{ \pstree{\TR{$etype(\asf,\emptyset)$}\trput{(3)}}
  {\pstree{\TR{$etype\_conj(\asf,\emptyset)$}~[tnpos=b]{$\land$}\trput{(4)}}
   {\pstree{\TR{$eseq(\epsilon,\{\asf\})$}\tlput{(5)/nil}}
    {\pstree{\TR{$eseq\_conj(\epsilon,\{\asf\})$}\trput{(6)}}
     {\TR{\false}\trput{(7)}
     }
    }
    {\psset{linestyle=dashed}
    \TR{$eseq(\ll\bsf,\asf\rl,\{\asf\})$}\trput{(5)/cons(,)}}
   }
  }
}
}
\end{minipage}
\begin{minipage}[t]{.3\textwidth}
\begin{tabbing}
XX \= XXXXX\kill
Legend:\\
\> $\asf =List(Even\tand\tneg Nat)$\\
\> $\bsf =Even\tand\tneg Nat$
\end{tabbing}
\end{minipage}

\caption{\label{fig:2} Evaluation of $etype(List(Even\tand\tneg
Nat))$}

\end{figure}

\endofproof
\end{example}

\begin{example} 
The following is a simplified version of the type definitions that
is used in~\cite{LuC:note} to show the incorrectness of the algorithm
by Dart and Zobel for testing inclusion of one regular type in
another~\cite{DartZ92}.

Let $\Pi=\{\alpha,\beta,\theta,\sigma,\omega,\zeta,\eta\}$,
$\Sigma=\{a,b,g(),h(,)\}$ and
\[ \rules = \left\{\begin{array}{llll}
               \alpha\rightarrow g(\omega), &
               \beta\rightarrow g(\theta)~|~g(\sigma),&
               \theta\rightarrow a~|~h(\theta,\zeta), &
               \sigma\rightarrow b~|~h(\sigma,\eta),\\
               \multicolumn{2}{l}{\omega\rightarrow a~|~b~|~h(\omega,\zeta)~|~h(\omega,\eta),} &
               \zeta\rightarrow{a}, & \eta\rightarrow{b}
            \end{array}\right\}   
\] \comments{Let
$\Sigma_{h}=\Sigma\setminus\{h\}$. We have
\begin{eqnarray*}
\sem{\rules}{\theta} &=& \{t\in\terms(\Sigma_{h})~|~\mbox{t is left-skewed and leaves of t are a's}\}\\
\sem{\rules}{\sigma} &=& \{t\in\terms(\Sigma_{h})~|~\mbox{t is left-skewed and leaves of t are b's}\}\\
\sem{\rules}{\omega} & = & \{t\in\terms(\Sigma_{h})~|~\mbox{t is left-skewed}\}\\
\sem{\rules}{\alpha} &=& \{g(t)~|~t\in\sem{\rules}{\omega}\}\\
\sem{\rules}{\beta} &=& \{g(t)~|~t\in\sem{\rules}{\theta}\cup\sem{\rules}{\sigma}\}
\end{eqnarray*}}

Let $t=g(h(h(a,b),a))$. $t\in\sem{\rules}{\alpha}$ and
$t\not\in\sem{\rules}{\beta}$, see example 3 in~\cite{LuC:note} for
more details.  So,
$\sem{\rules}{\alpha}\not\subseteq\sem{\rules}{\beta}$. This is
verified by our algorithm as follows. Let
$\Psi_{1}=\{\alpha\tand\tneg\beta\}$ and
$\Psi_{2}=\Psi_1\cup\{\omega\tand\tneg\theta\tand\tneg\sigma\}$.  By
applying equations~\ref{eq:top}, \ref{eq:rdo:1}, \ref{eq:rdo:2},
\ref{eq:rdt:1}, \ref{eq:rdt:2} and~\ref{eq:rdo:1} in that order, we
have
$etype(\alpha\tand\tneg\beta)=etype\_conj(\omega\tand\tneg\theta\tand\tneg\sigma,\Psi_1)$. By
equation~\ref{eq:rdo:2}, we have
\[etype(\alpha\tand\tneg\beta) = 
       eseq(\epsilon\tand\Lambda\tand\epsilon,\Psi_2)
~\land~  eseq(\epsilon\tand\epsilon\tand\Lambda,\Psi_2)
~\land~  eseq(\Theta,\Psi_2) 
\]
where $\Theta=(\ll\omega,\zeta\rl\tor\ll\omega,\eta\rl)
        \tand(\ll\tneg\theta,\ttop\rl\tor\ll\ttop,\tneg\zeta\rl)
        \tand(\ll\tneg\sigma,\ttop\rl\tor\ll\ttop,\tneg\eta\rl)$.  We
        choose not to simplify expressions such as
        $\epsilon\tand\epsilon\tand\tneg\Lambda$ so as to make the
        example easy to follow. By applying equations~\ref{eq:rdt:1}
        and~\ref{eq:rdt:2}, we have both
        $eseq(\epsilon\tand\Lambda\tand\epsilon,\Psi_2)=\true$ and
        $eseq(\epsilon\tand\epsilon\tand\Lambda,\Psi_2)=\true$. So,
        $etype(\alpha\tand\tneg\beta) = eseq(\Theta,\Psi_2)$.  Let
        $\Gamma=\ll\omega,\zeta\rl \tand\ll\tneg\theta,\ttop\rl
        \tand\ll\ttop,\tneg\eta\rl$. To show
        $etype(\alpha\tand\tneg\beta)=\false$, it suffices to show
        $eseq\_conj(\Gamma,\Psi_2)=\false$ by equation~\ref{eq:rdt:1}
        because $\Gamma\in{\dnf}(\Theta)$ and
        $etype(\alpha\tand\tneg\beta) =
        eseq(\Theta,\Psi_2)$. 

Figure~\ref{fig:3} depicts the evaluation of
$eseq\_conj(\Gamma,\Psi_2)$. The node that is linked to its parent by a
dashed line is not evaluated because one of its siblings returns
$\false$, which is sufficient to establish the falsity of its
parent. It is clear from the figure that
$etype\_conj(\Theta,\Psi_2)=\false$ and hence
$etype(\alpha\tand\tneg\beta)=\false$.

\begin{figure}
\psset{labelsep=2pt,tnpos=a,radius=2pt}
\pstree[nodesep=2pt,levelsep=25pt]
       {\TR{$etype\_conj(\Gamma,\Psi_2)$}~[tnpos=b]{$\lor$}}
{ \pstree{\TR{$etyp(\omega\tand\tneg\theta,\Psi_2)$}\tlput{7}}
  { \pstree{\TR{$etyp\_conj(\omega\tand\tneg\theta,\Psi_2)$}~[tnpos=b]{$\land$~~~~}\trput{4}}
    { \pstree{\TR{$eseq(\epsilon\tand\Lambda,\Psi_3)$}\tlput{5/a}}
      { \pstree{\TR{$eseq\_conj(\epsilon\tand\Lambda,\Psi_3)$}\tlput{6}}
        { \TR{\true}\tlput{7}
        }
      }
      \pstree{\TR{$eseq(\epsilon\tand\epsilon,\Psi_3)$}\trput{5/b}}
      { \pstree{\TR{$eseq\_conj(\epsilon\tand\epsilon,\Psi_3)$}\trput{6}}
        { \TR{\false}\trput{7}
        }
      }
      {\psset{linestyle=dashed}\TR{$eseq(\Theta_1,\Psi_3)$}\trput{5/h(,)}}
    }
  }
  \pstree{\TR{$etype(\zeta\tand\tneg\eta,\Psi_2)$}\trput{7}}
  { \pstree{\TR{$etype\_conj(\zeta\tand\tneg\eta,\Psi_2)$}\trput{4}}
    { \pstree{\TR{$eseq(\epsilon\tand\epsilon,\Psi_4)$}\trput{5}} 
      { \pstree{\TR{$eseq\_conj(\epsilon\tand\epsilon,\Psi_4)$}\trput{6}} 
        { \TR{\false}\trput{7} }
      }
    }
  }
}

\begin{tabbing}
XX \= XXXXX\kill
Legend:\\
\> $\Theta_1 = (\ll\omega,\zeta\rl\tor\ll\omega,\eta\rl)\tand
               (\ll\tneg\theta,\ttop\rl\tor\ll\ttop,\tneg\zeta\rl)$\\
\> $\Psi_3 =\Psi_2\cup\{\omega\tand\tneg\theta\}$\\
\> $\Psi_4 =\Psi_2\cup\{\zeta\tand\tneg\eta\}$\\
\> $\Gamma=\ll\omega,\zeta\rl \tand\ll\tneg\theta,\ttop\rl
        \tand\ll\ttop,\tneg\eta\rl$
\end{tabbing}

\caption{\label{fig:3} Evaluation of $etype\_conj(\Gamma,\Psi_2)$}

\end{figure}

\endofproof
\end{example}

\section{Correctness}\label{sec:soundness} 
This section addresses the correctness of the algorithm. We shall
first show that tabulation ensures the termination of the algorithm
because the table can only be of finite size. We then establish the
partial correctness of the algorithm.

\subsection{Termination} 
Given a type expression $E$, a top-level type atom in $E$ is a type
atom in $E$ that is not a sub-term of any type atom in $E$. The set of
top-level type atoms in $E$ is denoted by $\tlta{E}$. For instance,
letting $E=\tneg List(Nat)\tor Tree(Nat\tand\tneg Even)$,
$\tlta{E}=\{List(Nat), Tree(Nat\tand\tneg Even)\}$. We extend
$\tlta{\cdot}$ to sequences by $\tlta{{\ll
E_1,E_2,\cdots,E_k\rl}}\definedas\bigcup_{1\leq{i}\leq{k}}\tlta{E_{i}}$.

Given a type expression $E_{0}$, the evaluation tree for
$etype(E_{0})$ contains nodes of the form $etype(E,\Psi)$,
$etype\_conj(C,\Psi)$, $eseq(\Theta,\Psi)$ and
$eseq\_conj(\Gamma,\Psi)$ in addition to the root that is
$etype(E_{0})$. Only nodes of the form $etype\_conj(C,\Psi)$ add
conjunctive type expressions to the table. Other forms of nodes only
pass the table around. Therefore, it suffices to show that the type
atoms occurring in the first argument of the nodes are from a finite
set because any conjunctive type expression added into the table is
the first argument of a node of the form $etype\_conj(C,\Psi)$.

The set $\rta{E_{0}}$ of type atoms relevant to a type expression
$E_{0}$ is the smallest set of type atoms satisfying
\begin{itemize}
\item $\tlta{E_{0}}\subseteq\rta{E_{0}}$, and
\item if $\tau$ is in $\rta{E_{0}}$ and $\tau\rightarrow
         f(\tau_1,\tau_2,\cdots,\tau_k)$ is in ${ground}(\rules)$ then
         $\tlta{\tau_i}\subseteq\rta{E_{0}}$ for $1\leq{i}\leq{k}$.
\end{itemize} 
The height of $\tau_{i}$ is no more than that of $\tau$ for any
$\tau\rightarrow f(\tau_1,\tau_2,\cdots,\tau_k)$ in
${ground}(\rules)$. Thus, the height of any type atom in $\rta{E_{0}}$
is finite.  There are only a finite number of type constructors in
$\Pi$. Thus, $\rta{E_{0}}$ is of finite size. It follows by examining
the algorithm that type atoms in the first argument of the nodes in
the evaluation tree for $etype(E_{0})$ are from $\rta{E_{0}}$ which is
finite. Therefore, the algorithm terminates.

\subsection{Partial Correctness} 

The partial correctness of the algorithm is established by showing
$etype(E_{0})=\true$ iff $\tempty{E_{0}}$. Let $\Psi$ be a set of
conjunctive type expressions. Define $\rho_{\Psi}\definedas
\land_{C\in\Psi}\tempty{C}$. The following two lemmas form the core of
our proof of the partial correctness of the algorithm.

\begin{lemma}\label{lm:complete} 
Let $\Psi$ be a set of conjunctive type expressions, $E$ a type
expression, $C$ a conjunctive type expression, $\Theta$ a sequence
expression and $\Gamma$ a conjunctive sequence expression.
\begin{itemize} 
\item [(a)] If $\rho_{\Psi}\models\tempty{C}$ then
            $etype\_conj(C,\Psi)=\true$, and
\item [(b)] If $\rho_{\Psi}\models\tempty{E}$ then
            $etype(E,\Psi)=\true$, and
\item [(c)] If $\rho_{\Psi}\models\tempty{\Gamma}$ then
            $etype(\Gamma,\Psi)=\true$, and
\item [(d)] If $\rho_{\Psi}\models\tempty{\Theta}$ then
            $etype(\Theta,\Psi)=\true$.
\end{itemize}

\begin{proof} 
The proof is done by induction on the size of the complement of $\Psi$
with respect to the set of all possible conjunctive type expressions
in which type atoms are from $\rta{E_{0}}$ where $E_{0}$ is a type
expression.

Basis.  The complement is empty.  $\Psi$ contains all possible
        conjunctive type expressions in which type atoms are from
        $\rta{E_{0}}$. We have $C\in\Psi$ and hence
        $etype\_conj(C,\Psi)=\true$ by equation~\ref{eq:rdo:2}.
        Therefore, (a) holds. (b) follows from (a) and
        equation~\ref{eq:rdo:1}. (c) follows from (b),
        equation~\ref{eq:rdt:2} and lemma~\ref{lm:xxx} in the appendix,
        and (d) follows from (c) and equation~\ref{eq:rdt:1}.

Induction. By lemma~\ref{lm:x} in the appendix,
           $\rho_{\Psi}\models\tempty{C}$ implies $\rho_{\Psi}\models
           \tempty{\bcal_{C}^f}$ for any
           $f\in\cap_{\alpha\in{pos}(C)}\fcal(\alpha)$.  Thus,
           $\rho_{\Psi\cup\{C\}}\models \tempty{\bcal_{C}^f}$. The
           complement of $\Psi\cup\{C\}$ is smaller than the
           complement of $\Psi$. By the induction hypothesis, we have
           $eseq(\bcal_{C}^f,\Psi\cup\{C\})=\true$. By
           equation~\ref{eq:rdo:2},
           $etype\_conj(C,\Psi)=\true$. Therefore, (a) holds. (b)
           follows from (a) and equation~\ref{eq:rdo:1}. (c) follows
           from (b), equation~\ref{eq:rdt:2} and lemma~\ref{lm:xxx} in
           the appendix and (d) follows from (c) and
           equation~\ref{eq:rdt:1}. This completes the proof of the
           lemma.

\endofproof
\end{proof}
\end{lemma}

Lemma~\ref{lm:complete} establishes the completeness of $etype(,)$,
$etype\_conj(,)$, $eseq(,)$ and $eseq\_conj(,)$ while the following
lemma establishes their soundness.

\begin{lemma}\label{lm:sound} 
Let $\Psi$ be a set of conjunctive type expressions, $E$ a type
expression, $C$ a conjunctive type expression, $\Theta$ a sequence
expression and $\Gamma$ a conjunctive sequence expression.
\begin{itemize} 
\item [(a)] $\rho_{\Psi}\models\tempty{C}$ if
            $etype\_conj(C,\Psi)=\true$, and
\item [(b)] $\rho_{\Psi}\models\tempty{E}$ if
            $etype(E,\Psi)=\true$, and
\item [(c)] $\rho_{\Psi}\models\tempty{\Gamma}$ if
            $etype(\Gamma,\Psi)=\true$, and
\item [(d)] $\rho_{\Psi}\models\tempty{\Theta}$ if
            $etype(\Theta,\Psi)=\true$.
\end{itemize}

\begin{proof} 
It suffices to prove (a) since (b),(c) and (d) follow from (a) as in
lemma~\ref{lm:complete}.  The proof is done by induction on
$dp(C,\Psi)$ the depth of the evaluation tree for
$etype\_conj(C,\Psi)$.

Basis.  $dp(C,\Psi)=1$.  $etype\_conj(C,\Psi)=\true$ implies
        either (i) $pos(C)\cap{neg}(C)\neq\emptyset$ or (ii) $\exists
        C'\in\Psi.C\cleq C'$. In case (i), $\tempty{C}$ is true
        and $\rho_{\Psi}\models\tempty{C}$. Consider case (ii).  By
        the definition of $\cleq$ and $\rho_{\Psi}$, we have
        $etype\_conj(C,\Psi)=\true$ implies
        $\rho_{\Psi}\models\tempty{C}$.

Induction.  $dp(C,\Psi)>1$. Assume $etype\_conj(C,\Psi)=\true$ and
         $\rho_{\Psi}\models\neg\tempty{C}$. By lemma~\ref{lm:x}, there is
         $f\in\cap_{\alpha\in{pos}(C)}\fcal(\alpha)$ such that
         $\rho_{\Psi}\models\neg\tempty{\bcal_{C}^f}$. We have
         $\rho_{\Psi\cup\{C\}}\models\neg\tempty{\bcal_{C}^f}$. $dp(\bcal_{C}^f,\Psi\cup\{C\})<dp(C,\Psi)$. By
         the induction hypothesis, we have
         $etuple(\bcal_{C}^f,\Psi\cup\{C\})=\false$ for otherwise,
         $\rho_{\Psi\cup\{C\}}\models\bcal_{C}^f$. By
         equation~\ref{eq:rdo:2}, $etype\_conj(C,\Psi)=\false$ which
         contradicts $etype\_conj(C,\Psi)=\true$. So,
         $\rho_{\Psi}\models\tempty{C}$ if
         $etype\_conj(C,\Psi)=\true$. This completes the induction and
         the proof of the lemma.

\endofproof
\end{proof}
\end{lemma}

The following theorem is a corollary of lemmas~\ref{lm:complete}
and~\ref{lm:sound}.
\begin{theorem} For any type expression $E$, $etype(E)=\true$ 
                iff $\tempty{E}$. 

\begin{proof} 
By equation~\ref{eq:top}, $etype(E)=etype(E,\emptyset)$. By
lemma~\ref{lm:complete}.(b) and lemma~\ref{lm:sound}.(b), we have
$etype(E,\emptyset)=\true$ iff $\rho_{\emptyset}\models\tempty{E}$.
The result follows since $\rho_{\emptyset}=\true$.
\end{proof}
\end{theorem} 

\section{Complexity}\label{sec:complexity} 
We now address the issue of complexity of the algorithm.  We only
consider the worst-case time complexity of the algorithm.  The time
spent on evaluating $etype(E_{0})$ for a given type expression $E_{0}$
can be measured in terms of the number of nodes in the evaluation tree
for $etype(E_{0})$.

 The algorithm cycles through $etype(,),
etype\_conj(,), eseq(,)$ and $eseq\_conj(,)$. Thus, children of a node
of the form $etype(E,\Psi)$ can only be of the form
$etype\_conj(C,\Psi)$, and so on.

 Let $|S|$ be the number of elements in a given set $S$.  The largest
 possible table in the evaluation of $etype(E_{0})$ contains all the
 conjunctive type expressions of which type atoms are from
 $\rta{E_{0}}$. Therefore, the table can contain at most
 $2^{|\rta{E_{0}}|}$ conjunctive type expressions.  So, the height of
 the tree is bounded by $\ocal(2^{|\rta{E_{0}}|})$.

We now show that the branching factor of the tree is also bounded by
$\ocal(2^{|\rta{E_{0}}|})$.  By equation~\ref{eq:rdo:1}, the number of
children of $etype(E,\Psi)$ is bounded by two to the power of the
number of type atoms in $E$ which is bounded by $|\rta{E_{0}}|$
because $E$ can only contain type atoms from $\rta{E_{0}}$.  By
equation~\ref{eq:rdo:2}, the number of children of
$etype\_conj(C,\Psi)$ is bounded by $|\Sigma|$.  The largest number of
children of a node $eseq(\Theta,\Psi)$ is bounded by two to the power
of the number of sequences in $\Theta$ where $\Theta=\bcal_{C}^f$. For
each $\tau\in{neg}(C)$, $|push(\tneg(\tor\acal_{\tau}^f))|$ is
$\ocal(arity(f))$ and $|C|<|\rta{E_{0}}|$. Thus, the number of
sequences in $\Theta$ is $\ocal(arity(f)*|\rta{E_{0}}|)$ and hence the
number of children of $eseq(\Theta,\Psi)$ is
$\ocal(2^{|\rta{E_{0}}|})$ since $arity(f)$ is a constant. By
equation~\ref{eq:rdt:2}, the number of children of
$eseq\_conj(\Gamma,\Psi)$ is bounded by $\max_{f\in\Sigma}arity(f)$.
Therefore, the branching factor of the tree is bounded by
$\ocal(2^{|\rta{E_{0}}|})$.

The above discussion leads to the following conclusion.
\begin{proposition} The time complexity
of the algorithm is $\ocal(2^{|\rta{E_{0}}|}))$.

\endofproof
\end{proposition}

The fact that the algorithm is exponential in time is expected because
the complexity coincides with the complexity of deciding the emptiness
of any tree automaton constructed from the type expression and the
type definitions. A deterministic frontier-to-root tree automaton
recognising $\sem{\rules}{E_{0}}$ will consist of $2^{|\rta{E_{0}}|}$
states as observed in the proof of lemma~\ref{lm:regular}. It is
well-known that the decision of the emptiness of the language of a
deterministic frontier-to-root tree automaton takes time polynomial in
the number of the states of the tree automaton. Therefore, the
worst-case complexity of the algorithm is the best we can expect from
an algorithm for deciding the emptiness of regular types that contain
set operators.

\section{Conclusion}\label{sec:conclusion} 

We have presented an algorithm for deciding the emptiness of
prescriptive regular types.  Type expressions are constructed from
type constructors and set operators. Type definitions prescribe the
meaning of type expressions.

The algorithm uses tabulation to ensure termination. Though the
tabulation is inspired by Dart and Zobel~\cite{DartZ92}, the decision
problem we consider in this paper is more complex as type expressions
may contain set operators. For that reason, the algorithm can also be
used for inclusion and equivalence problems of regular types. The way
we use tabulation leads to a correct algorithm for regular types while
the Dart-Zobel algorithm has been proved incorrect for regular
types~\cite{LuC:note} in general. To the best of our knowledge, our
algorithm is the only correct algorithm for prescriptive regular
types.

In addition to correctness, our algorithm generalises the work of
Dart and Zobel~\cite{DartZ92} in that type expressions can
contain set operators and type definitions can be
parameterised. Parameterised type definitions are more natural than
monomorphic type
definitions~\cite{FruhwirthSVY:LICS91,Mycroft:OKeefe:84,Yardeni:Shapiro:91}
while set operators makes type expressions concise.  The combination
of these two features allows more natural type declarations. For
instance, the type of the logic program $append$ can be declared or
inferred as $append(List(\alpha),List(\beta),List(\alpha\tor\beta))$.

The algorithm is exponential in time. This coincides with deciding the
emptiness of the language recognised by a tree automaton constructed
from the type expression and the type definitions. However, the
algorithm avoids the construction of the tree automaton which cannot be
constructed {\it a priori} when type definitions are parameterised.

Another related field is set constraint
solving~\cite{AikenW92,AikenL94,heintze-set,heintze-decision,DevienneTT98}.
However, set constraint solving methods are intended to infer
descriptive types~\cite{Reddy:NACLP90} rather than for testing the
emptiness of a prescriptive type~\cite{Reddy:NACLP90}. Therefore, they
are useful in different settings from the algorithm presented in this
paper. In addition, algorithms proposed for solving set
constraints~\cite{AikenW92,AW93,AikenL94,AikenKVW93} are not
applicable to the emptiness problem we considered in this paper. Take
for example the {\em constructor rule} in~\cite{AikenW92,AikenL94}
which states that emptiness of $f(E_1,E_2,\cdots,E_m)$ is equivalent
to the emptiness of $E_{i}$ for some $1\leq{i}\leq{m}$. However,
$\tempty{List(\tbot)}$ is not equivalent to $\tempty{\tbot}$. The
latter is true while the former is false since
$\sem{\rules}{List(\tbot)}=\{nil\}$. The constructor rule doesn't
apply because it deals with function symbols only but doesn't take
the type definitions into account.


\appendix

\section*{Appendix}

\begin{lemma} \label{lm:x}  Let $C$ be a conjunctive
type expression.  $\tempty{C}$ iff
\[\begin{array}{l}
 \forall{f\in\cap_{\alpha\in{pos}(C)}\fcal(\alpha)}.\\
   \tempty{(\bigtand_{\omega\in{pos}(C)}(\tor \acal_{\omega}^f)) \tand
   (\bigtand_{\tau\in{neg}(C)}\tneg(\tor \acal_{\tau}^f)) } \end{array} 
\]

\begin{proof} Let $\bt$ be a sequence of terms and $f$ a function symbol. 
 By the definition of $\sem{\rules}{\cdot}$, 
 $f(\bt)\in\sem{\rules}{C}$ iff
 $f\in\cap_{\alpha\in{pos}(C)}\fcal(\alpha)$ and
 $\bt\in\sem{\rules}{\bigtand_{\omega\in{pos}(C)}(\tor
 \acal_{\omega}^f))}\setminus \sem{\rules}{
 (\bigtor_{\tau\in{neg}(C)}(\tor \acal_{\tau}^f))
 }$. $\bt\in\sem{\rules}{\bigtand_{\omega\in{pos}(C)}(\tor
 \acal_{\omega}^f))}\setminus \sem{\rules}{
 (\bigtor_{\tau\in{neg}(C)}(\tor \acal_{\tau}^f)) }$ iff
 $\bt\in\sem{\rules}{(\bigtand_{\omega\in{pos}(C)}(\tor
 \acal_{\omega}^f)) \tand (\bigtand_{\tau\in{neg}(C)}\tneg(\tor
 \acal_{\tau}^f))}$. Thus, $\tempty{C}$ iff
 $\tempty{(\bigtand_{\omega\in{pos}(C)}(\tor \acal_{\omega}^f)) \tand
 (\bigtand_{\tau\in{neg}(C)}\tneg(\tor \acal_{\tau}^f)) }$ for each
 $f\in\cap_{\alpha\in{pos}(C)}\fcal(\alpha)$.

\endofproof 
\end{proof}
\end{lemma}

\begin{lemma} \label{lm:xxx} 
Let $\Gamma$ be a conjunctive sequence expression. Then
\[\tempty{\Gamma}~~ \mbox{iff}~~\tor_{1\leq{j}\|\Gamma\|}
\tempty{\proj{\Gamma}{j}}\]

\begin{proof} Let $\|\Gamma\|=n$ and 
 $\Gamma=\gamma_1\tand\gamma_2\tand\cdots\tand\gamma_{m}$ with
 $\gamma_{i}=\ll\gamma_{i,1},\gamma_{i,2},\cdots,\gamma_{i,n}\rl$.  We
 have
 $\sem{\rules}{\Gamma}=\bigcap_{1\leq{j}\leq{m}}\sem{\rules}{\gamma_j}$.
 We have
 $\proj{\Gamma}{j}=\gamma_{1,j}\tand\gamma_{2,j}\tand\cdots\tand\gamma_{m,j}$.
 $\exists{1}\leq{j}\leq{n}.\tempty{\proj{\Gamma}{j}}$ iff
 $\exists{1}\leq{j}\leq{n}.\bigcap_{1\leq{i}\leq{m}}\sem{\rules}{\gamma_{i,j}}=\emptyset$
 iff $\sem{\rules}{\Gamma}=\emptyset$ iff $\tempty{\Gamma}$.

\endofproof
\end{proof} 
\end{lemma}

\begin{lemma} \label{lm:regular}
  $\sem{\rules}{\mcal}$ is a regular term language for any type expression $\mcal$.
\begin{proof} 
The proof is done by constructing a regular term grammar for
$\mcal$~\cite{GecsegS84}.
We first consider the case
$\mcal\in\terms(\Pi\cup\{\ttop,\tbot\})$. Let
$R=\ll\rta{\mcal},\Sigma,\emptyset,\Upsilon,\mcal\rl$ with
\[\Upsilon=\{(\alpha\rightarrow f(\alpha_1,\cdots,\alpha_{k}))
             \in{ground}(\rules)~|~\alpha\in\rta{\mcal}\}
\]
$R$ is a regular term grammar. It now suffices to prove that
$t\in\sem{\rules}{\mcal}$ iff $\mcal\Rightarrow_{R}^{*}t$.
\begin{itemize}
\item Sufficiency. Assume $\mcal\Rightarrow_{R}^{*}t$. The proof is
      done by induction on derivation steps in
      $\mcal\Rightarrow_{R}^{*}t$.  \begin{itemize} \item
      Basis. $\mcal\Rightarrow_{R}t$. $t$ must be a constant and
      $\mcal\rightarrow t$ is in $\Upsilon$ which implies
      $\mcal\rightarrow t$ is in $ground(\rules)$. By the definition
      of $\sem{\rules}{\cdot}$. $t\in\sem{\rules}{\mcal}$.  \item
      Induction. Suppose $\mcal\Rightarrow
      f(\mcal_1,\cdots,\mcal_{k})
      \Rightarrow_{R}^{(n-1)}t$. Then $t=f(t_1,\cdots,t_{k})$
      and $\mcal_i\Rightarrow_{R}^{n_{i}}t$ with $n_{i}\leq{(n-1)}$.  By
      the induction hypothesis, $t_i\in\sem{\rules}{\mcal_{i}}$ and
      hence $t\in\sem{\rules}{\mcal}$ by the definition of
      $\sem{\rules}{\cdot}$.  \end{itemize}
\item Necessity. Assume $t\in\sem{\rules}{\mcal}$. The proof is done
      by the height of $t$, denoted as $height(t)$.  \begin{itemize}
      \item $height(t)=0$ implies that $t$ is a constant.
      $t\in\sem{\rules}{\mcal}$ implies that $\mcal\rightarrow{t}$ is
      in $ground(\rules)$ and hence $\mcal\rightarrow{t}$ is in
      $\Upsilon$. Therefore, $\mcal\Rightarrow_{R}{t}$.  \item Let
      $height(t)=n$. Then $t=f(t_1,\cdots,t_{k})$.
      $t\in\sem{\rules}{\mcal}$ implies that $(\mcal\rightarrow
      f(\mcal_1,\cdots,\mcal_{k}))\in{ground}(\rules)$ and
      $t_i\in\sem{\rules}{\mcal_{i}}$.  By the definition of
      $\Upsilon$, we have $(\mcal\rightarrow
      f(\mcal_1,\cdots,\mcal_{k}))\in \Upsilon$. By the
      definition of $\rta{\cdot}$, we have
      $\mcal_{i}\in\rta{\mcal}$. By the induction hypothesis,
      $\mcal_{i}\Rightarrow_{R}^{*}t_{i}$. Therefore,
      $\mcal\Rightarrow_{R}f(\mcal_1,\cdots,\mcal_{k})
      \Rightarrow_{R}^{*}f(t_1,\cdots,t_{k})=t$.  \end{itemize}
\end{itemize}

Now consider the case
$\mcal\in\terms(\Pi\cup\{\tand,\tor,\tneg,\ttop,\tbot\})$. We complete
the proof by induction on the height of $\mcal$.
\begin{itemize}
\item $height(\mcal)=0$. Then $\mcal$ doesn't contain set operator. We
      have already proved that $\sem{\rules}{\mcal}$ is a regular term language.
\item Now suppose $height(\mcal)=n$. If $\mcal$ doesn't contain set
      operator then the lemma has already been proved. If the
      principal type constructor is one of set operators then the
      result follows immediately as regular term languages are closed under
      union, intersection and complement
      operators~\cite{GecsegS84,GecsegS96,ComonDGLTT98}. It now
      suffices to prove the case
      $\mcal=c(\mcal_1,\cdots,\mcal_{\l})$ with $c\in\Pi$.  Let
      $\ncal=c(X_1,\cdots,X_{\l})$ where each $X_{j}$ is a
      different new type constructor of arity $0$. 

      Let $\Pi'=\Pi\{X_1,\cdots,X_{\l}\}, \Sigma'=
           \Sigma\cup\{x_1,\cdots,x_{\l}\}$ and
           $\rules'=\rules\cup\{X_j\rightarrow
           x_j|1\leq{j}\leq{\l}\}$.  $\sem{\rules'}{\ncal}$ is a
           regular term language on
           $\Sigma\cup\{x_1,\cdots,x_{\l}\}$ because $\ncal$
           doesn't contain set operators. By the induction hypothesis,
           $\sem{\rules}{{\mcal_{j}}}$ is a regular term language. By
           the definition of $\sem{\cdot}{\cdot}$, we have
           \[\sem{\rules}{\mcal}=\sem{\rules'}{\ncal}[x_1:=\sem{\rules}{\mcal_1},\cdots,x_{\l}:=\sem{\rules}{\mcal_{\l}}]\]
           which is a regular term
           language~\cite{GecsegS84,GecsegS96,ComonDGLTT98}.
           $S[y_1:=S_{y_1},\cdots,]$ is the set of terms each of which
           is obtained from a term in $S$ by replacing each occurrence
           of $y_j$ with a (possibly different) term from $S_{y_j}$.
           This completes the induction and the proof.
\end{itemize}

The proof also indicates that a non-deterministic frontier-to-root
tree automaton that recognises $\sem{\rules}{\mcal}$ has
$|\rta{\mcal}|$ states and that a deterministic frontier-to-root tree
automaton that recognises $\sem{\rules}{\mcal}$ has
$\ocal(2^{|\rta{\mcal}|})$ states.

\endofproof
\end{proof}
\end{lemma}

\end{document}